\documentclass[10pt, conference]{IEEEtran}

\usepackage[T1]{fontenc}
\usepackage[utf8]{inputenc}

\usepackage[pdftex]{graphicx}
\usepackage{listings}
\usepackage{url}
\usepackage{pgfplots}
\usepackage{pgf-pie}
\pgfplotsset{compat=1.17}
\usepackage{booktabs}
\usepackage{dot2texi}
\usepackage{amssymb}
\usepackage{diagbox}
\usepackage{makecell}
\usepackage{amsmath}
\usepackage{mathtools}
\usepackage{multirow}
\allowdisplaybreaks

\definecolor{logging-blue}{RGB}{1, 130, 172}
\definecolor{dark-red}{rgb}{0.3,0.1,0.1}
\definecolor{dark-green}{rgb}{0.1,0.3,0.1}
\definecolor{dark-blue}{rgb}{0.1,0.1,0.5}
\usepackage[colorlinks,linkcolor=dark-red,citecolor=dark-green,urlcolor=dark-blue]{hyperref}
\usepackage[nameinlink]{cleveref}

\makeatletter
\let\@ORGmakecaption\@makecaption
\long\def\@makecaption#1#2{\@ORGmakecaption{#1}{#2}\vskip\belowcaptionskip\relax}
\makeatother

\usetikzlibrary{shapes, arrows, patterns, automata, positioning}

\newcommand\copyrighttext{%
  \footnotesize \textcopyright~2024 IEEE. Personal use of this material is permitted. Permission from IEEE must be obtained for all other uses, in any current or future media, including reprinting/republishing this material for advertising or promotional purposes, creating new collective works, for resale or redistribution to servers or lists, or reuse of any copyrighted component of this work in other works. Cite this article as follows: L. Sadlek, M. Hus\'{a}k, and P. \v{C}eleda. \textit{Identification of Device Dependencies Using Link Prediction}, NOMS 2024-2024 IEEE/IFIP Network Operations and Management Symposium, 2024, pp. 1-10, doi: \href{https://doi.org/10.1109/NOMS59830.2024.10575713}{10.1109/NOMS59830.2024.10575713}.}
\newcommand\copyrightnotice{%
\begin{tikzpicture}[remember picture,overlay]
\node[anchor=south,yshift=12pt] at (current page.south) {\fbox{\parbox{\dimexpr\textwidth-\fboxsep-\fboxrule\relax}{\copyrighttext}}};
\end{tikzpicture}%
}

\begin{document}

\title{Identification of Device Dependencies\\Using Link Prediction}

\author{\IEEEauthorblockN{Luk\'{a}\v{s} Sadlek\IEEEauthorrefmark{1}\IEEEauthorrefmark{2}, Martin Hus\'{a}k\IEEEauthorrefmark{1}, Pavel \v{C}eleda\IEEEauthorrefmark{2}}
\IEEEauthorblockA{\IEEEauthorrefmark{2}Faculty of Informatics, Masaryk University, Brno, Czech Republic\\}
\IEEEauthorblockA{\IEEEauthorrefmark{1}Institute of Computer Science, Masaryk University, Brno, Czech Republic\\
sadlek@mail.muni.cz, husakm@ics.muni.cz, celeda@fi.muni.cz}
}

\maketitle
\copyrightnotice

\begin{abstract}
Devices in computer networks cannot work without essential network services provided by a limited count of devices. Identification of device dependencies determines whether a pair of IP addresses is a dependency, i.e., the host with the first IP address is dependent on the second one. These dependencies cannot be identified manually in large and dynamically changing networks.
Nevertheless, they are important due to possible unexpected failures, performance issues, and cascading effects.
We address the identification of dependencies using a new approach based on graph-based machine learning. The approach belongs to link prediction based on a latent representation of the computer network's communication graph. It samples random walks over IP addresses that fulfill time conditions imposed on network dependencies. The constrained random walks are used by a neural network to construct IP address embedding, which is a space that contains IP addresses that often appear close together in the same communication chain (i.e., random walk). Dependency embedding is constructed by combining values for IP addresses from their embedding and used for training the resulting dependency classifier.
We evaluated the approach using IP flow datasets from a controlled environment and university campus network that contain evidence about dependencies. Evaluation concerning the correctness and relationship to other approaches shows that the approach achieves acceptable performance. It can simultaneously consider all types of dependencies and is applicable for batch processing in operational conditions.

\end{abstract}

\renewcommand\IEEEkeywordsname{Keywords}

\section{Introduction}

Each network contains devices that provide essential services, e.g., domain name service that translates domain names to IP addresses or domain controller that enforces active directory policy for Windows devices. Network communication reveals that essential services are often targeted with requests upon which other devices rely. The mapping of asset dependencies has a supportive function for other tasks in network management (e.g., reveals performance issues in practice) and is usable for risk analysis of critical systems~\cite{SolarWinds},~\cite{depMapping}.

Automated dependency detection has been studied for a long time since manually determining these dependencies is infeasible~\cite{Zand2015}. The motivation is often network configuration management, e.g., analysis of potential impacts in case of failures, performance issues, and malicious attacks~\cite{Chen2008},~\cite{Bahl2007},~\cite{SolarWinds}. The previous research revealed dependencies using passively monitored network traffic~\cite{natarajan2012}, active approach~\cite{Zand2014}, system logs instead of network traces~\cite{Lan2021}, and applying time series or graph mining~\cite{Lange2017time},~\cite{Slimani2020}. Even though it provided valuable results, it focused on specific input data, had some limitations, or revealed only specific types of dependencies.

The current research results from graph-based machine learning can improve the methods since machine learning is recommended for complex network topologies~\cite{faddom}. Link prediction was shown to be useful for recommender systems in social networks, spam detection, and network routing~\cite{Kumar2020}. Moreover, latent graph representation learning (alternatively node embedding) reveals the hidden structure of graphs. In other words, it transforms nodes to their low-dimensional embedding representation and can identify even relationships not captured in the input data. Hence, algorithms can be more efficient with it than with the original graph~\cite{Zhang2020}. 

We propose an approach that uses latent graph representation learning (inspired by the Node2Vec approach~\cite{Grover2016node2vec}) for a new use case, which is to compute dependency embedding for the training of a dependency classifier. We create its novel core and most complex part -- the custom exploration of communication chains present in data. We introduce conditions for timestamps of IP flows, which must be fulfilled by communication chains, discuss how to prepare input IP flows, and accomplish other design adjustments of graph representation learning. Our contribution also includes the measurement of the approach's properties on data from a controlled environment and campus network and its comparison with local similarity indices using correlation coefficients.

In this paper, we focus on two research questions:
\begin{enumerate}
    \item \textit{Can we identify device dependencies using graph-based machine learning for the link prediction problem?}
    \item \textit{What correctness, time aspects, dependency types, and amount of processed data of the link prediction approach for device dependency identification can we obtain?}
\end{enumerate}
We also focus on passively collected network traffic in the form of IP flows as input data for our approach.

This paper is organized as follows. Section~\ref{sec:rel_work} describes the related work from graph-based network analysis and defines device dependencies. Section~\ref{sec:method} proposes a novel method for device dependency identification using link prediction. Implementation of the method is introduced in Section~\ref{sec:implementation}, including conditions related to types of dependencies. Evaluation concerning the method's correctness, time aspects, dependency types, and amount of data is explained in Section~\ref{sec:evaluation}. The last Section~\ref{sec:conclusion} concludes the paper.

\section{Related Work} \label{sec:rel_work}

Graph-based approaches to data analysis are not uncommon in the network security domain. The attack graphs, for example, are the most well-known application of graphs in this domain. Their construction and usage were exhaustively covered by Kaynar et al.~\cite{Kaynar2016}. Akoglu et al.~\cite{Akoglu2014} surveyed graph-based techniques for anomaly detection in diverse domains, including network traffic analysis. The application of graphs in network-wide situational awareness was covered by Noel et al.~\cite{Noel2018} or recently by Husák et al.~\cite{Husak2023}. Bowman and Huang~\cite{Bowman2021} reviewed the challenges of the application of Graph AI in cyber security. Atzmüller and Kanawati~\cite{Atzmueller2022} provided an overview of explainability for complex network analysis in cyber security. Lagraa et al.~\cite{Lagraa2023} surveyed the application of graphs in intrusion and botnet detection.

\begin{figure}[b]
    \includegraphics[width=\columnwidth]{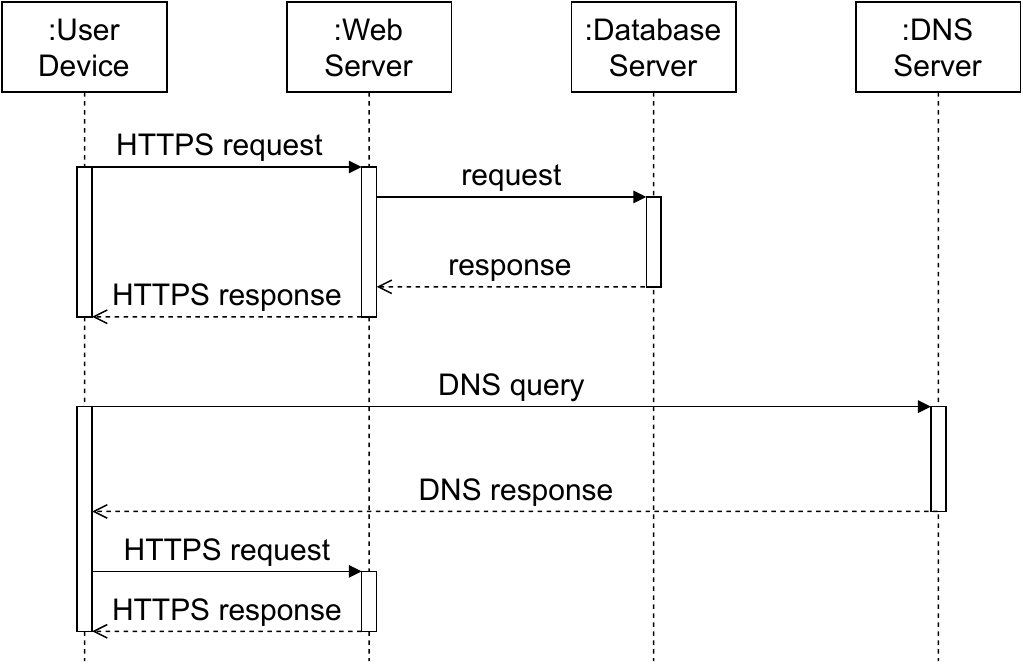}
    \caption{A sequence diagram containing local-remote dependency of the web server on the database server (the first activation of the user device) and remote-remote dependency of the web server on the DNS server (the second activation of the user device). Activations (vertical rectangles) denote participation of lifelines. Time passes from top to bottom in the diagram.} \label{fig:dependencies}
\end{figure}

A prime example of the application of advanced graph-based techniques in network security management is criticality and dependency detection, i.e., finding which devices in the network are the most important or how they depend on each other~\cite{natarajan2012, Zand2015}.  
In this work, we recognize three essential dependency classes. 
\emph{Direct dependency} (DD) is a network connection between a source and destination IP address that repeats more than threshold times with the same IP flow parameters except for timestamps. The \emph{local-remote} (LR) dependency from the view of a requesting server (e.g., the webserver in Figure~\ref{fig:dependencies}) consists of communication with another (remote) server (e.g., the database server) to answer the request from the user device~\cite{natarajan2012}. The \emph{remote-remote} (RR) dependency describes that one server is indirectly dependent on another to provide functionality for user devices~\cite{Chen2008}. Both servers provide remote services for the user device. For example, the user device in Figure~\ref{fig:dependencies} cannot access the webserver without obtaining its IP address from the DNS server.

Correct estimation of criticality and dependencies is vital for securing and hardening the networks but is hard to achieve in large networks due to the lack of detailed situational awareness and local knowledge~\cite{Lastovicka2017, Husak2023}. 
Passive network measurement, e.g., NetFlow~\cite{Hofstede2014}, is the most widely used for dependency detection and is briefly described later in this section. Natarajan et al.~\cite{natarajan2012} proposed the NSDMiner suite that determines LR dependencies based on passive network monitoring techniques. Zand et al.~\cite{Zand2015} proposed the automatic detection of critical services in the network based on finding cliques in the graphs of correlated services, i.e., services active at similar times. Laštovička and Čeleda~\cite{Lastovicka2017} proposed graph centrality-based approaches. Lange et al.~\cite{Lange2017time} used the time series-based analysis of network traffic to detect dependencies of network services. The topic became of utmost importance in cloud management, and thus, the SCoRMiner was proposed by Slimani et al.~\cite{Slimani2020} to detect dependencies between cloud services using both network traces and application-level data and graph mining.

Nevertheless, passive network-based approaches are not the sole approach to dependency detection. 
Zand et al.~\cite{Zand2014} proposed an active approach based on delay injection and implemented it in the Ripper tool. 
Lan et al.~\cite{Lan2021} proposed an approach based on system log analysis rather than network traffic, aiming at cloud environment and dependency discovery in microservices. 
Aksoy et al.~\cite{Aksoy2021} presented the most elaborated method of important IP address selection based on Laplacian centrality in communication graphs. 
Recently, Husák et al.~\cite{Husak2023} summarized the approaches to criticality estimation, connected them with graph-theoretical background, and briefly commented on their usability in network security management. 
An example of a practical take on dependency detection is the Application Discovery and Dependency Mapping (ADDM) from the Server \& Application Monitor (SAM) by SolarWinds~\cite{SolarWinds}, which combines manual data entry and automated discovery.

Link prediction~\cite{Kumar2020} is one of the most widely-used graph-theoretical approaches in data analysis, with a vast application potential in cybersecurity as well. The goal is to predict or disclose future or missing links between entities. Pope et al.~\cite{Pope2019} proposed the use of tailored link prediction heuristics in this domain, while Noel and Swarup~\cite{Noel2023} used dependency-based link prediction for learning microsegmentation policies. 
It is worth noting that we are not aware of any work that would use this particular combination of the problem of device dependency detection and the approach of link prediction.

The rise of machine learning (ML) also covered the graph-based data. Since traditional ML methods mostly take low-dimensional vectors as the inputs, there is a need to embed the graphs or nodes in the graphs into vectors. A well-known approach inspiring this work is the Node2Vec~\cite{Grover2016node2vec}. An alternative node representation for ML is the DeepWalk~\cite{Perozzi2014}. Readers are kindly referred to the survey by Zhang et al.~\cite{Zhang2020} for a comprehensive background and comparison. 
Nevertheless, there are classes of ML that allow the processing of graph structures directly. For example, graph neural networks (GNNs) are becoming increasingly popular even in network security management, such as in intrusion detection~\cite{Lo2022}, malware detection~\cite{Busch2021}, or network slicing in digital twins~\cite{Wang2022}.

Since this and most of the related work relies on network flows or similar data, we briefly introduce them here as well. Flow monitoring is a prevalent network traffic monitoring in large-scale and high-speed networks~\cite{Hofstede2014}. Network flow is defined as a unidirectional sequence of packets that share the source and destination IP address, IP protocol number, and TCP or UDP port or ICMP code. The flows are exported in NetFlow or IPFIX formats and accompanied by further information, such as timestamp, duration, and number of transferred packets and bytes. A typical bidirectional network connection exports as two flows but can be later paired into biflows, special flow records describing network traffic in both directions. 
Readers are kindly referred to an exhaustive tutorial by Hofstede et al.~\cite{Hofstede2014}.


\section{Method for Identification of Dependencies} \label{sec:method}
\begin{figure}[b]
        \includegraphics[width=\hsize]{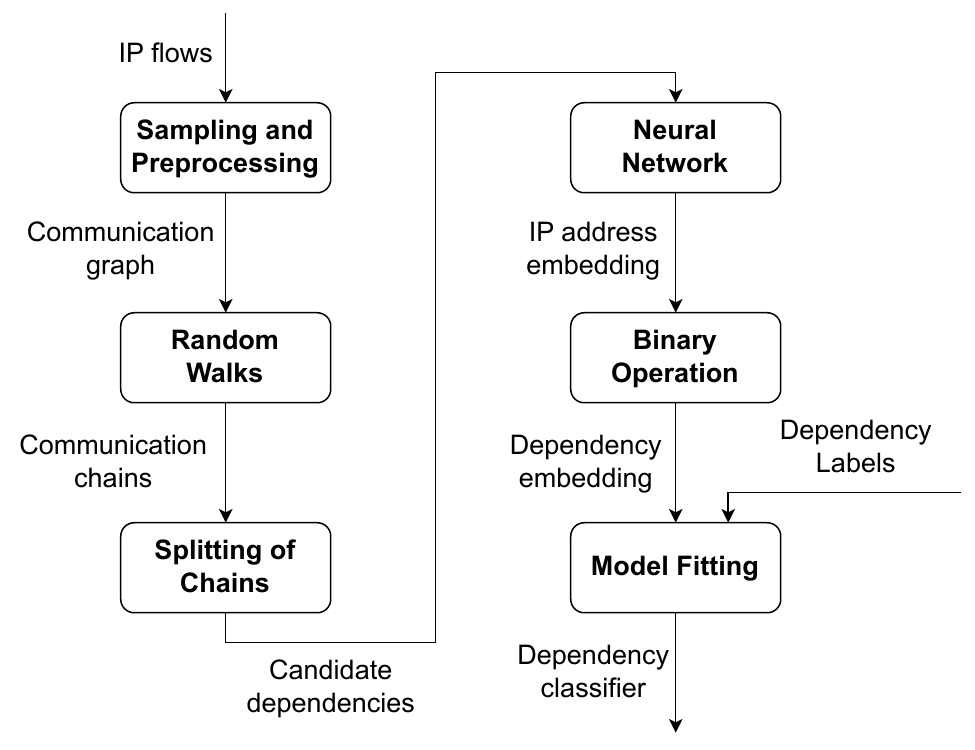}
    \caption{Steps of the proposed approach from processing of input data to obtaining dependency classifier.} \label{fig:approach}
\end{figure}

The proposed method for the identification of device dependencies creates a dependency classifier obtained via multiple steps expressed in Figure~\ref{fig:approach}.
Even though the approach processes bidirectional network communication, input IP flows must be unidirectional since we sample communication chains from a directed graph. Bidirectional IP flows must be converted to unidirectional using either distinct or the same start and end timestamps for both unidirectional flows.

The input IP flows can contain more IP addresses than can be processed by a neural network in a feasible time. It is necessary to focus only on IP addresses, among which we attempt to determine dependencies. We assume that the most important IP addresses have the highest number of IP flows except for attacks that produce a lot of communication, e.g., network scanning. When these IP addresses become vertices in a graph sample, we need a realistic communication neighborhood (i.e., communicants) for each vertex during random walk exploration. For this purpose, we should process enough edges, and sampling must be fair, i.e., all sampled edges must be chosen with equal probability. 

The requirement can be achieved by sampling with a reservoir of length $n$~\cite{vitter1985}. Any incoming $k$-th edge (i.e., IP flow for a specific IP address) receives a random number from $0$ to $k - 1$. If the number represents an index from the reservoir (i.e., $k < n$), the item is stored at such a position. As a result, we should obtain a directed communication multigraph that contains the source port, destination port, protocol, and start and end timestamps for each edge.

The next step is the core and the most complex one. It creates directed random walks representing communication chains of IP addresses. The chains are constrained by conditions imposed on timestamps of two subsequent IP flows to choose the next vertex of the random walk according to the current and previous vertex. 
Imposing conditions on more than two consequent IP flows in a communication chain could result in the trying of all possible sequences of higher length and drastically impact performance.

The first condition for creating communication chains is called the \textit{opening of LR dependency}, expressed as Condition~\ref{eq:1} in Section~\ref{sec:implementation}. It corresponds to a sequence of communication from the user device to the database server via the web server in Figure~\ref{fig:dependencies}. The second one is a \textit{return from LR dependency} related to a sequence from the database server to the user device via the web server in Figure~\ref{fig:dependencies}. It is mathematically described as Condition~\ref{eq:2} in Section~\ref{sec:implementation}. 

\textit{Opening of RR dependency} expresses a communication chain from the user device to the DNS server, from the DNS server to the user device, and then from the user device to the web server in Figure~\ref{fig:dependencies}. However, we omit redundant repeating user device from the communication chain because the DNS server and the web server would still be very close to it in the communication chain. Therefore, the final chain is the user device, the DNS server, and the web server. The opening of RR dependency is formally described by Condition~\ref{eq:3} in Section~\ref{sec:implementation}. The last condition is the \textit{return to the previous IP address} expressed as Condition~\ref{eq:4} that is fulfilled, e.g., by a communication chain of the web server, the database server, and the web server in Figure~\ref{fig:dependencies}.

The four conditions represent the building blocks of the communication chains. Chains that fulfill them can contain device dependencies in a form suitable for splitting in the next step. As a result, the approach also reveals transitive dependencies (TDs) composed of at least two DDs fulfilling conditions on LR dependency but with communication from solely one user device. An example of such a transitive dependency could be the dependency of the user device on the database server in Figure~\ref{fig:dependencies}, where also LR dependency exists between the web server and the database server.

The random walk exploration uses these conditions to create the inputted number of random walks starting in each vertex. The length of the random walk is also specified as input and should typically range from three to seven to capture dependencies. The second vertex of the random walk is chosen randomly from the vertices fulfilling the input threshold imposed on the number of IP flows, in which the first and second vertices appear as source and destination IP addresses (threshold $n_t$ in Table~\ref{tab:ground_truth}). The third and consequent vertices are randomly chosen out of vertices that fulfill at least one condition.
If no vertex fulfills a condition, then a vertex that fulfills the IP flow threshold is randomly chosen. If no vertex has enough appearances, then any target of directed edges from the source vertex is randomly chosen so that the random walk continues using graph edges.

The algorithm creates communication chains where directed edges are transformed from the captured communication in concordance with the order of communication steps required by the dependency types.  
If an IP address appears commonly in the same time-constrained communication chain with another IP address, then it is probable that one of them could be dependent on the other one. The longer the distance between the IP addresses in the chain, the lower the chance that dependency exists. Thus, it is beneficial to process the close pairs as candidate dependencies. 

\begin{figure}[t]
    \centering
    \includegraphics[width=0.95\columnwidth]{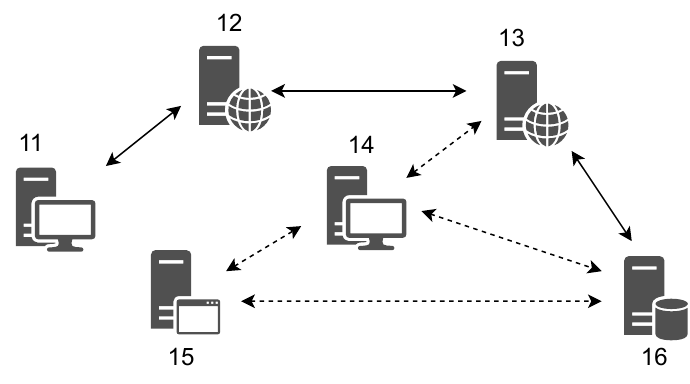} 
    \caption{Example of network communication between workstations and servers. Solid lines represent a communication chain. All edges consist of forward and reverse IP flows. Numbers are the last octets from IPv4 addresses.}
    \label{fig:splitting}
\end{figure}

Candidate dependencies are determined from the communication chain of a specific length using a sliding window (context), which we explain based on Figure~\ref{fig:splitting}. For simplicity, assume that a communication chain with length four consisting of $11$, $12$, $13$, and $16$ fulfills conditions. If the context size is three, then we obtain two contexts from this chain. The first is $11$, $12$, and $13$, and the second is $12$, $13$, and $16$. The candidate dependencies are pairs of addresses from one context where the first address is the initial address from the context, and the second one can be any of the following. We obtain candidate dependencies $(11, 12)$ and $(11, 13)$ from the first context and $(12, 13)$ and $(12, 16)$ from the second one.

The neural network uses the candidate dependencies to estimate features of the hidden layer so that IP addresses from candidate dependencies would be close together in the IP address embedding space, as is usual for embedding~\cite{Zhang2020}. IP address embedding contains a vector with a number of values equal to the number of features in the hidden layer for each IP address and must be transformed into dependency embedding using vectors for individual IP addresses. 

Binary operations such as average and L1 distance can transform node embedding to edge embedding~\cite{Grover2016node2vec}. Since the main purpose of embedding is to have similar nodes close together and distance is commutative, these operations should be commutative. In our case, we use a scalar product of two vectors to combine representation from node embedding to dependency embedding. It causes the range of values appearing in vectors to increase when we apply the product. In other words, it extends the space where candidate dependencies are mapped, and this space is then used for classification. As a result, the dependency embedding contains a vector of values for each candidate pair.

As a last step, each possible dependency in the embedding is given its label that denotes whether it is a dependency. Training the model with dependency embedding and labels (see Figure~\ref{fig:approach}) creates a dependency classifier that can determine whether a pair of IP addresses can be a dependency. This approach reveals the existence of the dependency, but it cannot automatically distinguish the direction of dependency because it uses embedding. Moreover, the revealed dependencies are influenced by the position of the IP flow collector. If we capture communication only at the edge of the network, then all dependencies containing one communication pair will be between one internal and one external IP address since the communication graph, in this case, will be bipartite between internal and external IP addresses. 

The method can provide as good results as its input IP flows. Tunneling approaches and anonymization protocols can replace original IP headers, and the approach will process the replaced IP addresses. However, a dependency on a Virtual Private Network (VPN) server that could be identified is valid when the VPN is needed to access internal network resources. Using anonymization protocols in the long term for communication with critical devices (such as DNS servers and cloud storage) is a rare scenario. The method also focuses only on the most essential IP flow properties from packet headers. Therefore, privacy-preserving protocols (e.g., HTTPS) that encrypt packet data do not hinder its use.

\section{Implementation of the Method} \label{sec:implementation}
The implementation consists of six steps, as depicted in Figure~\ref{fig:approach}. The first three are discussed separately, while the remaining are described in Subsection~\ref{subsec:emb}. Our implementation reuses existing supportive functionality of the Node2Vec approach in Python (mainly related to neural network training)~\cite{pyg_documentation}. However, we provide the method's most complex step of exploring valid random walks and implement the remaining parts to process IP flows. The implementation is available in supplementary materials~\cite{Sadlek2024supplementary}.

\subsection{Sampling and Data Preprocessing}
The first step obtains a representative sample with $n$ internal and $m$ external IPv4 addresses based on the highest number of IP flows from a batch of data. Consequently, we sample $k$ edges for the selected internal and external IP addresses. For this purpose, we implemented reservoir sampling that selects each edge with equal probability~\cite{vitter1985}.
Data preprocessing also includes
removing communication that does not use TCP and UDP protocols since, for them, the dependencies are defined in related work~\cite{natarajan2012} used for ground truth comparison in Section~\ref{sec:evaluation}.

\subsection{Random Walks} 
A directed random walk is a sequence $v_{1}, v_{2}, \ldots, v_{n}$ where vertex $v_{i+2}$ for $i \in \{1, \ldots, n-2\}$ is determined based on vertices $v_{i+1}$ and $v_{i}$. Since general node embedding approaches cannot directly support computer network graphs, it considers time constraints and IP flow properties, e.g., transport ports.

Let $t_1(v_{i}, v_{i+1})$ denote the start timestamp and $t_2(v_{i}, v_{i+1})$ the end timestamp of IP flow between vertices (i.e., IP addresses) $v_{i}$ and $v_{i+1}$. At least one of four conditions must hold for three subsequent vertices $v_{i}$, $v_{i+1}$, and $v_{i+2}$. The condition expressing opening of LR dependency: 
\begin{multline} \label{eq:1}
    t_1(v_{i}, v_{i+1}) \leq t_1(v_{i+1}, v_{i+2}) \leq t_2(v_{i+1}, v_{i+2}) \leq \\ \leq t_2(v_{i}, v_{i+1}), i \in \{1, \ldots, n-2\}
\end{multline}
holds for the forward direction of LR dependency. It expresses that the first request from a user device to the server (e.g., the web server in Figure~\ref{fig:dependencies}) is followed by another request from that server to another server (e.g., the database server in Figure~\ref{fig:dependencies}) to process the original request. Therefore, the first IP flow between the user device and the server will not end until the additional request to another server is processed.

The condition for return from LR dependency:
\begin{multline} \label{eq:2}
    t_1(v_{i+1}, v_{i+2}) \leq t_1(v_{i}, v_{i+1}) \leq t_2(v_{i}, v_{i+1}) \leq \\ \leq t_2(v_{i+1}, v_{i+2}) \land \exists j: j \neq i \land v_{i+2} = v_{j} \land v_{i+1} = v_{j+1}, \\ i \in \{1, \ldots, n-2\}, j \in \{1, \ldots, n-1\}
\end{multline}
describes the reverse direction of LR dependency (i.e., a chain of reverse IP flows), assuming that the random walk already contains its forward direction. Consider LR dependency from Figure~\ref{fig:dependencies}. In this case, $v_{i}$ denotes the database server, $v_{i+1}$ the web server, and $v_{i+2}$ the user device. The IP flow from the web server to the user device starts first, but the IP flow from the database server to the web server will end first. The second condition differs from the first one in the order of vertices.

The third condition: 
\begin{multline} \label{eq:3}
    t_2(v_{i}, v_{i+1}) \leq t_1(v_{i}, v_{i+2}) \land \\ \land t_1(v_{i}, v_{i+2}) - t_2(v_{i}, v_{i+1}) \leq \varepsilon, i \in \{1, \ldots, n-2\}
\end{multline}
expresses the opening of RR dependency that should happen within the specified time $\varepsilon$. For example, let $v_{i}$ be the user device from Figure~\ref{fig:dependencies}, $v_{i+1}$ the DNS server, and $v_{i+2}$ the web server. The forward IP flow from the user device to the DNS server, represented as edge $(v_{i}, v_{i+1})$, and its reverse flow that ends approximately at the same time will end before the request to the web server. The third condition may also include sequences that accidentally fulfill the condition within $\varepsilon$, but they will be less frequent per time unit with longer observation and smaller input parameter $\varepsilon$.

The fourth condition for triplets of form $v_{1}$, $v_{2}$, and $v_{1}$ formalizes return over reverse edge representing reverse flow. In this case, two unidirectional flows are represented by edges $(v_{1}, v_{2})$ and $(v_{2}, v_{1})$, where the former is the initiator flow.
\begin{multline} \label{eq:4}
    s(v_{1}, v_{2}) = d(v_{2}, v_{1}) \land d(v_{1}, v_{2}) = s(v_{2}, v_{1}) \land \\ \land t_1(v_{1}, v_{2}) \leq t_1(v_{2}, v_{1}) \land \\ \land \lvert t_2(v_{1}, v_{2}) - t_2(v_{2}, v_{1}) \rvert \leq \varepsilon, i \in \{1, \ldots, n-1\}
\end{multline}
expresses that the source ($s$) and destination ($d$) ports are equal but mutually exchanged, and we use time constraints for their starts and ends. 

While the brute-force approach would check that the dependency materialized at least $n_t$ times in data, the random walk exploration checks that there are at least $n_t$ sampled IP flows between the last processed and the next vertex that should extend the random walk. Consequently, all vertices fulfilling one of the conditions are chosen with equal probability.

\subsection{Splitting of Chains}
Communication chains (i.e., random walks) obtained from the previous step should be divided into candidate dependencies (i.e., pairs of IP addresses) representing input for the neural network.
For this purpose, we consider the sliding windows of the neighboring IP addresses in the communication chain, also called context in DeepWalk and Node2Vec~\cite{Perozzi2014, Grover2016node2vec}. We use the one-sided context for the proposed approach based on the following argumentation, contrary to the general link prediction on undirected graphs. 

Consider Condition~\ref{eq:1} holding for triplet $v_i$, $v_{i+1}$, and $v_{i+2}$ from forward random walk.  
Without loss of generality, assume that $(v_{i}, v_{i+1})$ and $(v_{i+1}, v_{i+2})$ are forward IP flows. If we consider the double-sided window, then we also consider reverse random walk, i.e., edges $(v_{i+2}, v_{i+1}), (v_{i+1}, v_{i})$, to the learning process as a valid sequence of IP addresses.
In the optimal case, the forward IP flow has a lower start timestamp than the reverse IP flow, and they end approximately at the same time.
Therefore, we obtain $ t_1(v_{i}, v_{i+1}) \leq t_1(v_{i+1}, v_{i}) \leq t_2(v_{i+1}, v_{i}) \approx t_2(v_{i}, v_{i+1})$ and similarly $t_1(v_{i+1}, v_{i+2}) \leq t_1(v_{i+2}, v_{i+1}) \leq t_2(v_{i+2}, v_{i+1}) \approx t_2(v_{i+1}, v_{i+2})$. 

We know that $ t_1(v_{i}, v_{i+1}) \leq t_1(v_{i+1}, v_{i+2}) \leq t_2(v_{i+1}, v_{i+2}) \leq t_2(v_{i}, v_{i+1})$ holds for forward edges from Condition~\ref{eq:1}. Denote $A = t_1(v_{i+1}, v_{i})$, $B = t_2(v_{i+1}, v_{i})$, $C = t_1(v_{i+2}, v_{i+1})$, and $D = t_2(v_{i+2}, v_{i+1})$. One possibility of ordering $A$, $B$, $C$, and $D$ using $\leq$ operator and assuming approximate equations with $\approx$ is $CADB$ that does not correspond to any of Conditions~\ref{eq:1} --~\ref{eq:4}.
It means that we would often incorrectly consider the opposite direction of dependency as a valid dependency for neural network training.

\subsection{Embedding and Model Fitting} \label{subsec:emb}
The context from the previous step is processed by the neural network in Python. All candidate dependencies $(v_i, v_j)$ for the same first vertex $v_i$ are processed simultaneously, not separately, because the neural network can adjust the embedding so that the first vertex is close to all specified vertices.
The neural network also uses stochastic gradient descent, a skip-gram model, and negative sampling optimization to obtain IP address embedding~\cite{Grover2016node2vec}. 

We use positive dependency labels (ground truth) and an equal number of negative labels for random pairs of vertices to compute the embedding of their source and destination IP addresses. Consequently, the scalar product of these values creates an item for an IP address pair in dependency embedding. We use a random forest classifier in our implementation to fit dependency embedding to the dependency labels.

\section{Evaluation} \label{sec:evaluation}
In this section, we describe the evaluation datasets and the ground truth. Correctness, time aspects, and other properties of the method were evaluated using a proof-of-concept implementation based on PyTorch Geometric~\cite{Fey2019fast},~\cite{pyg_documentation}. Moreover, we compare the proposed approach with local similarity indices and discuss lessons learned. The evaluation was accomplished using Python implementation on a personal computer with 64 GB RAM, 16 CPU cores, and a processor's clock speed of 2.5 GHz. All parameters of evaluation are available in supplementary materials~\cite{Sadlek2024supplementary}.

\subsection{Datasets}
We used two kinds of IPFIX flow datasets captured in network topologies familiar to us in detail. The first kind was captured during a two-day cyber defense exercise~\cite{tovarnak2020data, tovarnakzenodo}. The cyber exercise involved six teams (denoted as T1 -- T6) that had the same emulated network but behaved in a different way, which provides six partial datasets. Moreover, the exercise used one global network common for all teams. Bidirectional IP flows were captured at the edge of team networks. We converted them to unidirectional form prior to the evaluation. We did not consider IP addresses that represented attacker machines.

The second type was captured in the university campus network, which is assigned class B address space with /16 CIDR prefix. Two partial campus datasets contain ten-minute-long (denoted as U10m) and one-hour-long (denoted as U1h) time windows captured at the network edge during working hours on one Tuesday in March 2022 (U10m) and on one Wednesday in February 2023 (U1h). The campus network data contained unidirectional IP flows.

\subsection{Ground Truth}
We determined the ground truth by brute-forcing all possible sequences of unidirectional IP flows up to a limited length of four vertices, which is sufficient because of the context size we consider. Table~\ref{tab:ground_truth} contains the number of direct dependencies (DD) and RR dependencies with two (RR) or three communication pairs (RR3) for considered datasets. The dependencies appeared at least $n_t$ times. Two subsequent requests in one RR dependency were accomplished within $\varepsilon$. Many RR dependencies were DNS dependencies.

We do not distinguish LR dependencies because they are already present as DDs. However, we list transitive dependencies that can be created by chaining two (TD) or three direct dependencies (TD3) that fulfill LR conditions on timestamps. It is as if a specific user device solely caused the materialization of LR dependency in the data. Hence, we add its dependency on the supportive server, e.g., the database server in Figure~\ref{fig:dependencies}. Duplicate TDs are allowed only when two distinct paths materialize the dependency.

\begin{table}[t]
    \caption{Thresholds ($n_t$, $\varepsilon$) and number of dependencies (DD, RR, RR3, TD, TD3) for team networks from cyber exercise (T1 -- T6) and datasets from campus network (U10m, U1h). U1h contains average values from twelve time windows and used different $n_t$ thresholds for DD and TD dependencies (50) and RR dependencies (300).}
    \label{tab:ground_truth}
    \centering
    \begin{tabular}{c||c|c|c|c|c|c|c|c}
        \toprule
         & \textbf{T1} & \textbf{T2} & \textbf{T3} & \textbf{T4} & \textbf{T5} & \textbf{T6} & \textbf{U10m} & \textbf{U1h} \\ \midrule
         $\boldsymbol{n_t}$ & 10 & 10 & 10 & 10 & 10 & 10 & 10 & 50/300\\
         $\boldsymbol{\varepsilon}$ & 1 s & 1 s & 1 s & 1 s & 1 s & 1 s & 0.5 s & 0.5 s \\ \midrule
        \textbf{DD} & 300 & 444 & 330 & 427 & 321 & 143 & 38,372 & 2,866 \\
        \textbf{RR} & 248 & 131 & 177 & 310 & 98 & 35 & 1,731 & 164 \\
        \textbf{RR3} & 1,875 & 113 & 697 & 2,067 & 161 & 26 & 23,905 & 2,347 \\
        \textbf{TD} & 17 & 13 & 22 & 24 & 9 & 14 & 854 & 117 \\
        \textbf{TD3} & 0 & 0 & 2 & 1 & 0 & 0 & 359 & 81 \\ \bottomrule
    \end{tabular}
\end{table}

We compared the ground truth with the results from the open-source implementation of NSDMiner~\cite{Nsdminer2015}. NSDMiner used the number of bidirectional IP flows listed in Table~\ref{tab:nsdminer}. We did not use transport ports from its output and limited the number of required appearances of dependencies to ten. Otherwise, we used the default options of NSDMiner.

Table~\ref{tab:nsdminer} shows that almost all dependencies with non-zero confidence and approximately half of all dependencies by NSDMiner were revealed. It indicates the validity of the ground truth since NSDMiner uses a different approach. It compares only the overall timestamps of biflows, while we compared these timestamps for forward and reverse IP flows. 

\begin{table}[t]
    \centering
    \caption{Count of LR dependencies for team networks (T1 -- T6) by NSDMiner and dependencies found in the ground truth. $C$ denotes confidence.}
    \begin{tabular}{c|c|c|c|c|c|c}
    \toprule
        & \textbf{T1} & \textbf{T2} & \textbf{T3} & \textbf{T4} & \textbf{T5} & \textbf{T6} \\ \midrule
        IP Flows (thousands) & 55.1 & 47.8 & 29.6 & 55.6 & 40.5 & 23.2 \\ \midrule
        Dependencies for $C > 0$ & 7 & 5 & 0 & 2 & 1 & 0 \\
        Present in ground truth & 6 & 5 & 0 & 2 & 1 & 0 \\\midrule
        All dependencies & 46 & 50 & 29 & 57 & 36 & 15 \\
        Present in ground truth & 27 & 18 & 17 & 28 & 20 & 11 \\ \bottomrule
    \end{tabular}
    \label{tab:nsdminer}
\end{table}

\subsection{Properties of the Method}
We evaluated the method's properties by exploring random walks with five vertices while the context size was four. The approach accomplished ten walks for each vertex, and each positive walk was balanced by finding one non-existing (i.e., negative) random walk. 
We used five learning epochs for training the model because only small changes in the computed loss between positive and negative random walks were observed after them. 
The number of values in the embedding vectors was 64, i.e., the embedding had 64 dimensions.

\begin{table*}[t]
    \centering
    \caption{Amount of data and measured time for team networks (T1 -- T6) and data from university campus network (U10m, U1h) using the proposed approach. U1h contains averages from twelve time windows except for IP flows and addresses.}
    \begin{tabular}{c|c|c|c|c|c|c|c|c|c}
    \toprule
        & & \textbf{T1} & \textbf{T2} & \textbf{T3} & \textbf{T4} & \textbf{T5} & \textbf{T6} & \textbf{U10m} & \textbf{U1h} \\ \midrule
        \multirow{4}*{Data} & IP flows & 61,346 & 54,941 & 34,721 & 63,266 & 46,253 & 28,506 & 8,259,584 & 78,270,416 \\ 
        & IP addresses & 689 & 1,421 & 654 & 1,190 & 1,047 & 247 & 451,365 & 1,235,300 \\
        & Vertices & 111 & 96 & 99 & 102 & 95 & 103 & 129 & 93 \\
        & Edges (contains multiple edges) & 21,026 & 21,720 & 20,035 & 22,905 & 20,986 & 16,080 & 15,076 & 18,411 \\ \midrule
        \multirow{3}*{Time} & Preprocessing & 12.80 s & 14.88 s & 8.41 s & 13.86 s & 11.54 s & 6.19 s & 27.15 s & 27.32 s \\
        & Creating embedding & 15.93 min & 14.02 min & 14.06 min & 14.83 min & 12.98 min & 13.95 min & 6.23 min & 6.93 min \\ 
        & Computation & $<$ 3 s & $<$ 1 s & $<$ 2 s & $<$ 2 s & $<$ 1 s & $<$ 3 s & $<$ 1 s & $<$ 1 s \\ \bottomrule
    \end{tabular}
    \label{tab:properties}
\end{table*}

\begin{table*}[t]
    \caption{Evaluation metrics for individual teams (T1 -- T6) from cyber defense exercise and two campus network datasets (U10m, U1h). Accuracy, precision, and F1 score contain averages from fifteen train-test splits except for U1h that was also split into twelve consequent windows.}
    \label{tab:metrics}
    \centering
    \begin{tabular}{c|c||c|c|c|c|c|c|c|c}
        \toprule
        Test size & & \textbf{T1} & \textbf{T2} & \textbf{T3} & \textbf{T4} & \textbf{T5} & \textbf{T6} & \textbf{U10m} & \textbf{U1h} \\ \midrule
        0.25 & Accuracy & 0.514 & 0.417 & 0.493 & 0.503 & 0.429 & 0.337 & 0.479 & 0.515 \\
        & Precision & 0.612 & 0.521 & 0.597 & 0.605 & 0.530 & 0.444 & 0.604 & 0.615 \\
        & F1 score & 0.666 & 0.566 & 0.644 & 0.656 & 0.572 & 0.462 & 0.625 & 0.664 \\ \midrule
        0.50 & Accuracy & 0.535 & 0.453 & 0.529 & 0.532 & 0.485 & 0.403 & 0.515 & 0.545 \\
        & Precision & 0.628 & 0.544 & 0.621 & 0.630 & 0.580 & 0.485 & 0.612 & 0.638 \\
        & F1 score & 0.677 & 0.584 & 0.669 & 0.672 & 0.617 & 0.510 & 0.645 & 0.681 \\ \midrule
        --- & AUC & 0.68 & 0.64 & 0.67 & 0.68 & 0.67 & 0.61 & 0.71 & 0.69 \\
        & AP & 0.81 & 0.74 & 0.80 & 0.81 & 0.78 & 0.68 & 0.84 & 0.81 \\ \bottomrule
    \end{tabular}
\end{table*}

We used the test sizes of 25\% and 50\% of all labels for individual train-test splits. The number of positive and negative labels was the same, but we did not constrain which were chosen into the test set. All datasets represented one time window except for a one-hour-long dataset from the university campus network that was divided into twelve consequent (i.e., five-minute-long) time windows since such time windows can be obtained from IP flow collectors in practice. The method was executed on each window separately, i.e., nineteen times.

Datasets for correctness evaluation contained numbers of IP flows and IP addresses listed in Table~\ref{tab:properties}. A sampling of input IP flows outputted approximately one hundred of the most important IP addresses according to the number of IP flows in which they participated.
Creating embedding (i.e., neural network training) took approximately a quarter of an hour, mainly due to the exploration of constrained random walks. A long learning time is usual for neural network approaches that infer complex relationships.
Computation for considered test sizes was executed in one to three seconds and took a much shorter time than brute-forcing all possibilities.

\begin{figure}[t]
  \centering

  \includegraphics[width=\columnwidth]{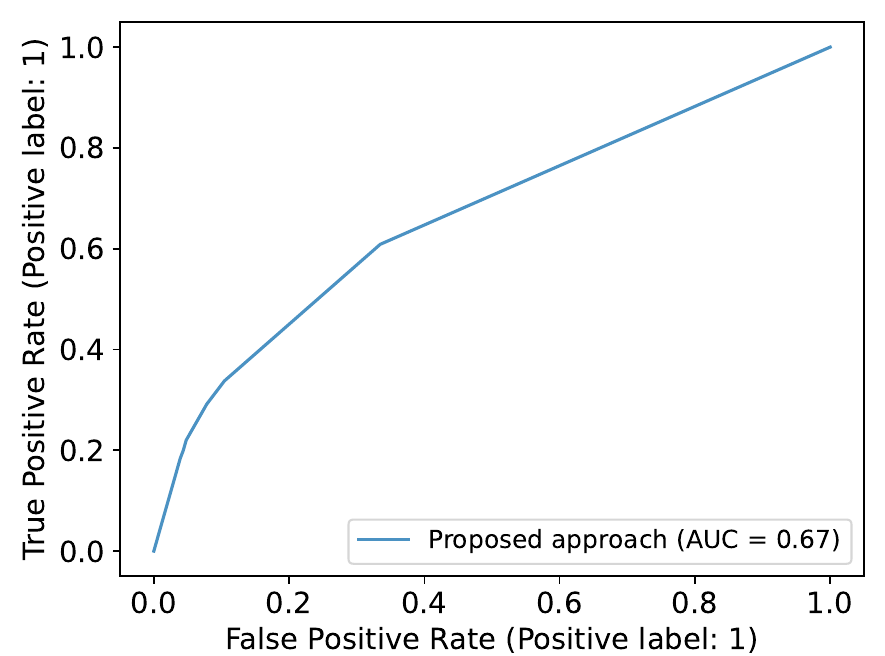}
  \newline
  \newline
  \includegraphics[width=\columnwidth]{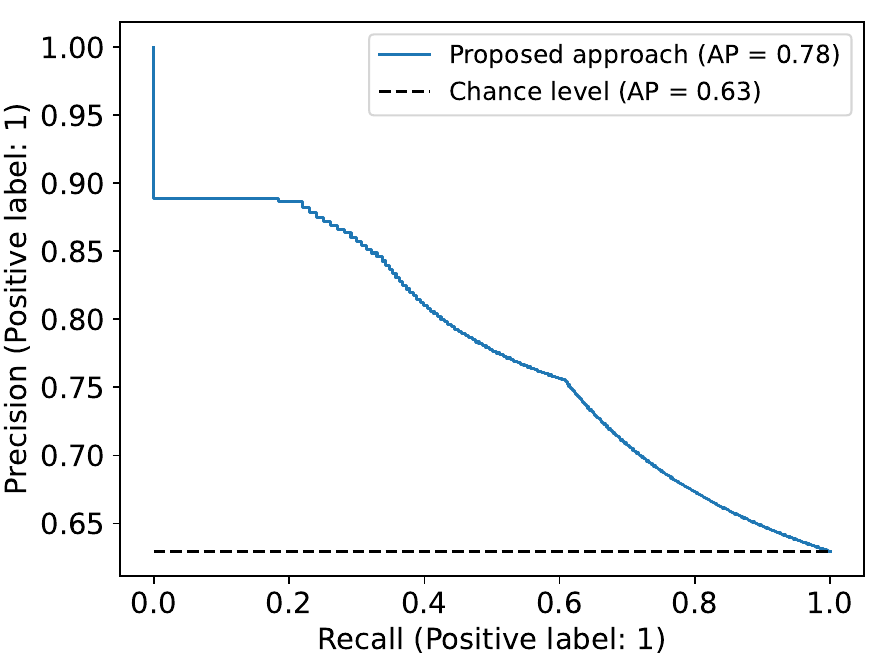}

  \caption{The ROC and PR curves of the proposed approach for team five from the cyber defense exercise.} \label{fig:cyber_czech}
\end{figure}

Table~\ref{tab:metrics} contains accuracy, precision, F1 score, area under receiver operating characteristic (ROC) curve (denoted as AUC), and average precision (AP) for the datasets. Since the F1 score is a harmonic mean of precision and recall, it is a suitable measure for imbalanced datasets where most labels belong to one class. AP summarizes precision-recall (PR) curves. The curves for team networks are available in supplementary materials~\cite{Sadlek2024supplementary}. Examples for team five are depicted in Figure~\ref{fig:cyber_czech}. The perfect ROC curve is very close to the upper left, and the PR curve to the upper right corner.

The proposed approach obtained approximately as good AUC, AP, and other metrics for small datasets from cyber defense exercise as for university campus network data. The average AUC for all windows from the one-hour-long campus network dataset was 0.69, while all values ranged between 0.63 and 0.74 AUC. The average AP for the twelve windows was 0.81, and the APs ranged from 0.74 to 0.88.

The accuracy, precision, and F1 score are not as comprehensive as AUC and AP because test sizes of 25\% and 50\% may not be fractions where they achieve their optimal values. Yang et al.~\cite{Yang2015evaluating} recommend using AP (i.e., the area under the PR curve) since it can better cope with a typical class imbalance of link prediction. The random classifier has an AUC of 0.5, but its AP equals the fraction of positive labels among all labels. The chance level is listed in all PR curves in supplementary materials for six teams~\cite{Sadlek2024supplementary}. It ranged from 61\% to 73\% for university data, i.e., the majority class was the positive one. 
Due to these facts, we conclude that the approach provides acceptable performance with respect to the class imbalance.

\subsection{Comparison with Local Similarity Indices} \label{subsec:sim}
The method should determine more complex dependencies, not only DDs or missing edges, even though the random walks were sampled over IP flows, which represent graph edges and possible DDs. Positive results from correctness evaluation could mean that it focused too much on DDs 
(i.e., existing edges with many occurrences).

Local similarity indices provide a substantially different approach because they use paths containing three vertices and determine only possible edges in the graph. 
We used four local similarity indices -- Adamic-Adar (AA), Common Neighbors (CN), Preferential Attachment (PA), and Resource Allocation (RA). We adjusted them to directed variants:
\begin{align*}
    s_{AA}(x, y) & = \sum_{v \in N_{out}(x) \cap N_{in}(y) : \lvert N_{out}(v) \rvert \neq 1} \frac{1}{log \lvert N_{out}(v) \rvert} \\
    s_{CN}(x, y) & = \lvert N_{out}(x) \cap N_{in}(y) \rvert \\
    s_{PA}(x, y) & = \lvert N_{out}(x) \rvert \cdot \lvert N_{in}(y) \rvert \\
    s_{RA}(x, y) & = \sum_{v \in N_{out}(x) \cap N_{in}(y)} \frac{1}{\lvert N_{out}(v) \rvert}
\end{align*}
where $N_{out}(x)$ denotes a set of vertices $v$ connected from vertex $x$ by an edge $(x, v)$ and $ N_{in}(y)$ a set of vertices $v$ that are connected to $y$ by an edge $(v, y)$. These indices are equal to zero for distant nodes and are higher when they are closer.

The comparison of local similarity indices on data from cyber defense exercise with the results of our method determined correlation using the Spearman correlation coefficient and Kendall tau~\cite{yu2019analysis}.
Both were ranging from 0 to 11\%. It means that each local similarity index outputted uncorrelated values to the method's predicted values and also to probabilities assigned to pairs of IP addresses. It indicates that even though the method uses graph edges as input, the conditions on timestamps of IP flows cause it to determine more complex dependencies that are not directly visible in the input.

\subsection{Lessons Learned}
The proposed approach had 0.61 to 0.74 AUC for directed graphs when predicting dependencies using an imbalanced set of labels due to counting with all possible pairs of IP addresses.
A more general Node2Vec approach had 0.77 to 0.97 AUC for undirected graphs when only edges were predicted~\cite{Grover2016node2vec}. Since the task of link prediction for undirected graphs is much easier and its evaluation used an equal number of positive and negative labels, the measured AUCs for the proposed approach seem to be realistic.

The proposed approach can deal with multiple types of dependencies. However, the separate identification of LR dependencies may not be suitable for it due to the lack of labels (see Table~\ref{tab:nsdminer}).
Further tuning of the model can use a different classifier than random forest, and the parameters can be tuned for IP flow data using optimization methods, e.g., grid search and random search~\cite{yang2020hyperparameter}.

\section{Conclusion} \label{sec:conclusion}
This paper addressed device dependency identification based on passively collected IP flows. We used latent graph representation learning for a new use case of creating dependency embedding used by a dependency classifier. The novel core and the most complex part is based on creating communication chains fulfilling time conditions imposed on device dependencies. 

The proposed device dependency identification achieves acceptable correctness and is suitable for use cases when the training time does not represent a disadvantage. However, the prediction of dependencies using an already trained model can be very fast compared to using brute force approaches. It can cope with all dependency types simultaneously and process large amounts of data split into batches. 

Further research can couple the approach to device criticality detection, network topology discovery, and enterprise mission modeling. It can also be extended by results from active monitoring and adjusted to discover dependencies of network services deployed on stable ports and complex dependencies with redundant devices, such as alternative DNS servers.

The supplementary materials contain a proof-of-concept implementation, a detailed list of parameters used for evaluation, and plots drawn for data from the controlled environment~\cite{Sadlek2024supplementary}. All materials can be used to reproduce evaluation results.

\section*{Acknowledgment}

This research was supported by project ``MSCA\-fellow5\_MUNI'' (No. CZ.02.01.01/00/22\_010/0003229). 

\bibliographystyle{IEEEtran}
\bibliography{bibliography}

\end{document}